# An Agent-based Architecture for a Knowledge-work Support System

Arijit Laha
Center for Knowledge-driven
Information Systems (C-KDIS),
SETLabs, Infosys Technologies Ltd.
Hyderabad, India
arijit_laha@infosys.com

## ABSTRACT
Enhancement of technology-based system support for knowledge workers is an issue of great importance. The "Knowledge work Support System (KwSS)" framework analyzes this issue from a holistic perspective. KwSS proposes a set of design principles for building a comprehensive IT-based support system, which enhances the capability of a human agent for performing a set of complex and interrelated knowledge-works relevant to one or more target task-types within a domain of professional activities. In this paper, we propose a high-level, software-agent based architecture for realizing a KwSS system that incorporates these design principles. Here we focus on developing a number of crucial enabling components of the architecture, including (1) an Activity Theory-based novel modeling technique for knowledge-intensive activities; (2) a graph theoretic formalism for representing these models in a knowledge base in conjunction with relevant entity taxonomies/ontologies; and (3) an algorithm for reasoning, using the knowledge base, about various aspects of possible supports for activities at performance-time.

## 1. INTRODUCTION
Correct and efficient performance of a "professional knowledge work" (henceforth referred simply as "knowledge-work") can bring about wide-ranging benefits across multiple levels, ranging from individual to societal. A knowledge-work, also called a "task" or "project", is typically performed within a domain of professional activities, e.g., business, governance, basic and applied research, healthcare etc., in order to solve problems of tactical and/or strategic natures. Laha [9] proposed a framework for designing Information Warehouse (IW) as a specialized repository of granular and richly contextualized information sharable among a number of task-specific Knowledge-work Support Systems (KwSS). In this paper we are interested in the architectural aspects of a single KwSS. Thus, without loss of generality, we shall refer to the entire framework as the KwSS framework, of which the IW is a component that serves as a dedicated information repository.

The importance and relevance of the problem addressed by KwSS framework [9] can be seen from the recent spate of works and initiatives that are addressing various facets of the problem. The ASAP [4] and the Codex [14] attempt at significantly improving the support level for works in domains of genome research and geography research respectively. The US government's SHARP [13] project mandates research into issues for building comprehensive support systems for patient-care tasks. Also, there are a few ongoing research projects that attempt to enhance the level of support for knowledge-works in various domains. Examples of such initiatives include NEPOMUK - The Social Semantic Desktop (http://nepomuk.semanticdesktop.org/), X-Media (Large Scale Knowledge Sharing and Reuse across Media - http://www.x-media-project.org/), PALETTE (Pedagogically sustained Adaptive Learning through the Exploitation of Tacit and Explicit Knowledge - http://palette.ercim.org/).

Each of the above is designed for one or more predetermined tasks in a particular domain and address a limited number of facets of the problem space. In contrast, the KwSS framework is largely domain-agnostic and task-neutral, that can be leveraged to implement a KwSS system for any chosen task in any chosen domain. Further, the framework is based on a more holistic and deeper view of the problem than any of the above efforts. In the current paper, a general system architecture to aid/guide implementations of KwSS systems is developed. Here we use the notion of software agents [15] as components of the architecture because, (1) we envision a KwSS as an evolving system; and (2) we want different constituent modules/sub-systems to respond both on-demand as well as proactively.

## 2. SYSTEM SUPPORT: THE KWSS WAY
In the context of KwSS, a knowledge-work is performed by one or more knowledge-workers who possess the requisite expertise and experience. During the performance, a worker typically needs to gather a significant body of information from various sources, understand and interpret the information in the context of the current problem. This leads to the worker gaining knowledge about possible solution(s). Once in possession of the knowledge, the worker articulates it in the form of various sharable/communicable informational artifacts (plan, design, report, advise, etc.). In all, a knowledge-work is a complex interaction between human mind and environment aided by information and tools for manipulating information.

A KwSS is an information processing system and makes no claim at being able to do the "thinking" on behalf of a knowledge-worker. From functional viewpoint, the KwSS perceives a knowledge-work as *consumption/absorption of information* and *production/creation of new information*, together referred here as "information usage", by human agents. Essentially, a KwSS aims to create a support environment that significantly enhances the capability of a worker to find relevant (i.e., worth consuming) information as well as articulate and record new information worth communicating/sharing. One of the crucial and differentiating premises of the KwSS framework is that actual processes of information usage take place during performance of various smaller, *cognitively manageable* "knowledge-intensive activities", hereafter referred simply as "activities", which constitute a larger task. Based on these facts and their various implications [8, 9], the framework argues that a system, in order to significantly enhance support for performing episodes of a knowledge-work, must aim supporting granularity levels of these cognitively manageable activities, while maintaining the structure of the whole task.

To illustrate various points over the rest of the paper we shall use as examples, situations from *patient-care* as a knowledge-work, where a physician and her colleagues treat an ailing patient. We shall assume that a patient-care KwSS is being used for fulfilling requirements of information access, creation and recording. Note that, patient-care is chosen as an example as most readers are likely to be familiar with it. Both the framework and architecture developed later can be used for building system to support any knowledge-work. In the following subsection major aspects of information usage considered within the KwSS framework are described.

## 2.1 Cognition-related Support

Performance of a knowledge-work makes great demand on the cognitive/intellectual faculties of a knowledge-worker. Unfortunately, our cognitive ability to focus our attention to a task is innately limited [1]. To overcome this limitation, a common practice is to decompose a large, complex activity into smaller sub-activities until, given the availability of relevant resources (expertise, information, support systems), each of the granules of activities is cognitively manageable.

For example, the task of patient-care is divided into examination, diagnosis, treatment, follow-up and so on. The activity examination is further decomposed into recording of symptoms (headache, stiffness of limbs, etc.), finding/measuring signs (body temperature, blood pressure level, etc.) and collection of medical history. The KwSS design framework recommends that the support for a task should be extended to the level of granularity of activities, where they are actually performed, i.e., information is consumed and new knowledge gained and articulated. The granular level supports envisaged in KwSS include the areas described below.

### *2.1.1 Maintenance of Context*

In order to perform an activity, a knowledge-worker needs to construct and actively maintain a mental model of the work-context. This requires a high degree of cognitive effort. A KwSS strives to provide significant aid in this respect. It attempts to locate enough contextual in formation, present them to a worker and maintain as well as transfer it among interrelated activities so that the a human worker can use the information as cues/hints to (re)construct and effectively maintain her mental model with significantly lesser effort. In other words, a KwSS needs to maintain an adequate representation of the context of and across the activities it supports.

### *2.1.2 Access to relevant information*

During the performance of a knowledge-work, a worker needs to access significant volume of relevant information. The processes involved in information seeking and retrieval, judging their relevance and subsequent internalization by the human worker are highly complex ones [6]. Nonetheless, these processes are heavily influenced by the work-context. A KwSS attempts to use available contextual cues in order to support these processes at the granular activity level. It attempts to go beyond conventional document-level access and provide context-aware access to relevant information granules at text passage levels. Such an approach, along with providing more efficient access to information, also plays an important role in avoiding possible *information overload*.

### *2.1.3 Granular information articulation and capture*

Accessing and understanding relevant information allows a worker gain new knowledge/insight with respect to the problem-at-hand. As mentioned earlier, this actually happens when the worker is engaged in a cognitively manageable granule of activity. Naturally, this is the point of time when the knowledge and its context is most vivid in the worker's mind. Many details get lost with the passage of time. A KwSS supports a worker to articulate this knowledge without much delay, i.e., as part of performing the granular activity, as well as without significant additional effort. In other words, it should provide adequate means to produc, contextualize and capture the granular information efficiently.

## 2.2 Support for Behavioral issues

A KwSS is aimed to be used by a community of knowledge-workers, (sometimes known as a Community-of-Practice (CoP)). In such environments, several interesting issues arise which may impede usability and acceptance of such systems. Drawing upon the analysis by Markus et al. [11], the KwSS framework recommends system supports covering following areas.

### *2.2.1 Guidance*

A knowledge-work is performed by a human actor who possesses adequate expertise and experience. However, in real world not all workers can be expected to possess equal/similar level of expertise. Thus, a KwSS includes means to guide a user through a sequence of activities that is likely to result in a performance of (at least) acceptable quality.

### *2.2.2 Learning*

Performance of a knowledge-work itself is a major source of learning, often called the "on-the-job learning" for a knowledge worker. Such learning allows her to gain experience as well as to avoid professional obsolescence. Satisfying this need requires catering to a vast and varied information requirement, spanning across episodes of past performances, semantic and typological information, legal and various policy/practice-related information, information from relevant professional literature and many more. A KwSS is designed to provide efficient access to sources of such information.

### *2.2.3 Discretion or Autonomy of a User*

A knowledge-work is usually performed in order to solve a complex and often ill-structured, problem. There is no preferred or best structure for such an activity that can guarantee high quality outcomes. Knowledge-workers vary in expertise levels as well as in their preferred styles of reasoning. For example, Patel et al. in chapter 30 of [5] distinguish between hypothesis-driven and data-driven reasoning styles in medicine. Also, due to changes in environment, the problem may present novel, unprecedented features. Tackling them requires a worker to exercise her ingenuity. A KwSS allows practice of ingenuity by allowing ample scope to its users to exercise their “discretion” at performance-time.

# 3. ACTIVITY MODELING

To fulfill above support requirements, the system must be provided with enough actionable information about activities, information and their interdependency forming a basis for rich and fine-grained context-aware computing. This, in turn, requires that the system be equipped with a formal, i.e., machine-deployable, model of the supported activities that can represent the required information. To the best of our knowledge, none of the existing/established modeling techniques fits the bill. Our studies revealed that the Workflow-based techniques [16], while successful in modeling transactional and operational processes, cannot accommodate the complexity of knowledge-intensive activities. On the other hand, many Task-Analysis techniques [3] can be used for analyzing complex activities but are considerably difficult to formalize. Therefore, we develop a new formal “activity modeling” approach by co-opting some ideas from “Activity Theory (AT)” [12].

Before we proceed further, here we specify/reiterate a few terms and their semantics in context of the discussions ahead. A KwSS [9] system is designed to provide comprehensive support for performing “episodes” of a set of interrelated knowledge-intensive activities or knowledge-works in a domain of professional activities. In a KwSS, performance of each activity is identified within the span of performance of a larger unit of knowledge-work, called a “task”. In other words, a KwSS is designed to support at least one particular “task-type”, that will be refered as the “target-type” or “target” of that particular KwSS. For example, in a patient-care KwSS the target task-type is treatment of a patient, i.e., bringing an ailing person back to the state of health. Each episode/instance of the target, known as a “case” in the medical domain vocabulary, is performed with respect to the treatment of a particular patient, spanning from her admission to her discharge. The case, in turn, is a complex web of inter-dependent knowledge-intensive activities, e.g., examination, diagnosis, etc., each of which, in itself, is a complex knowledge-work.

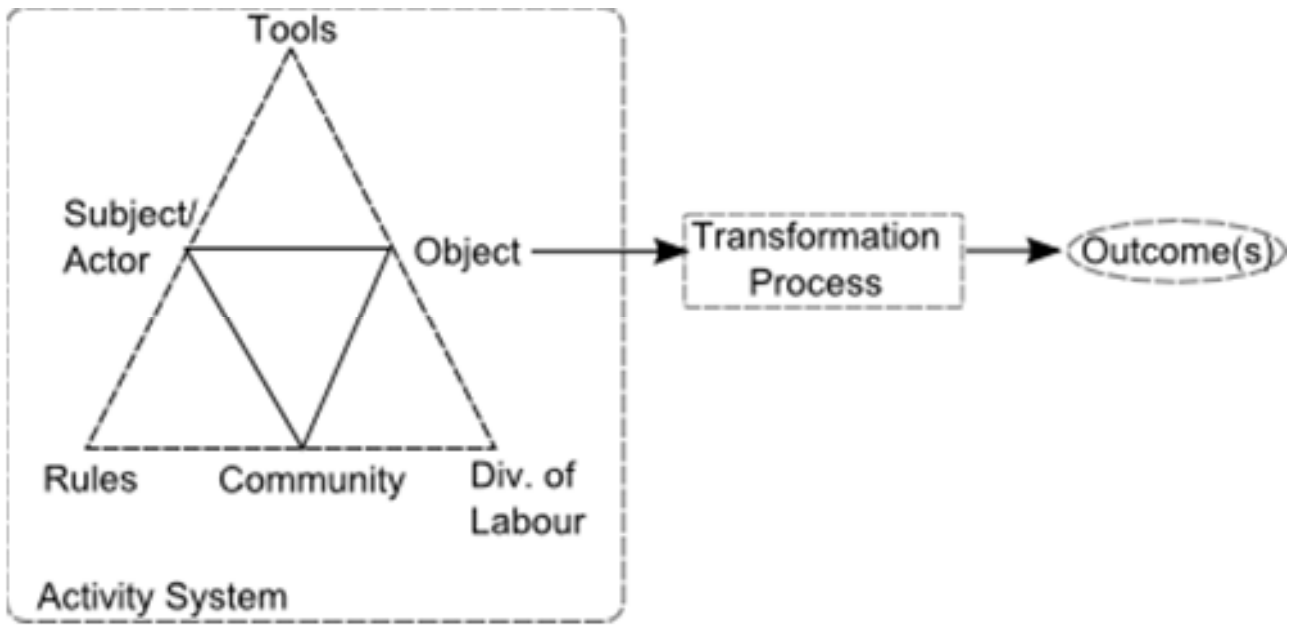

**Figure 1: Model of a general “Human Activity” in AT**

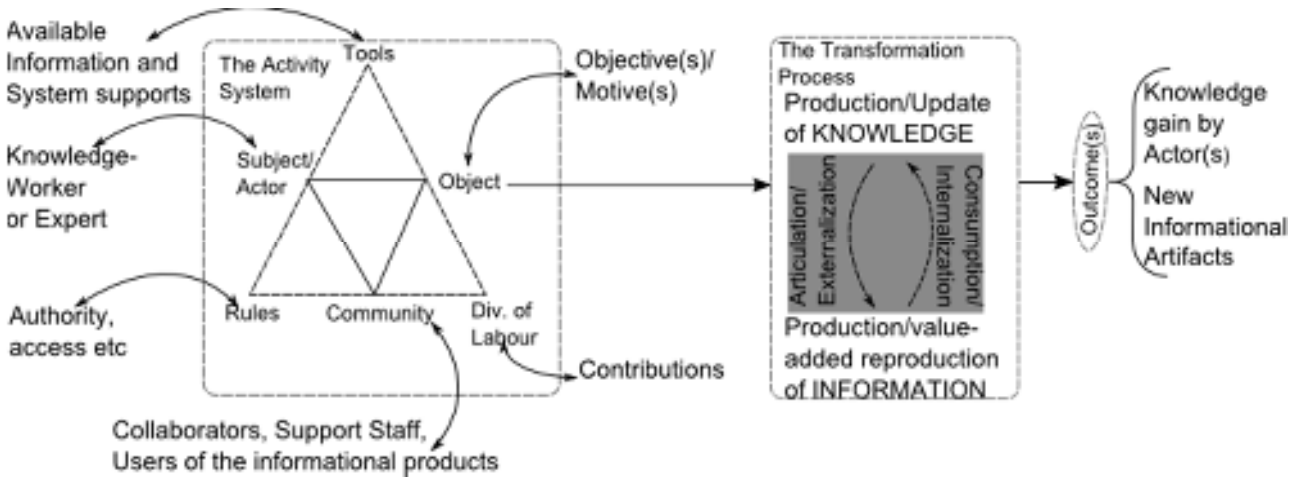

**Figure 2: Knowledge-work in light of AT**

## 3.1 Model of Activity in AT

The notion of a knowledge-work can be highly complex. Here we co-opt the general model of “human activity” from Activity Theory (AT) [7] as shown in Figure 1. According to AT, an activity is essentially an interaction between a subject or human actor and an object mediated by a set of tools. The element “object” covers two different senses, firstly, some entity (physical or abstract) that is manipulated or transformed (including from state of non-existence to existence) in the course of activity, and secondly, the objective(s)/motives of the activity. “Tools” refer to concrete (e.g., a machine), mental (e.g., expertise, experience) and informational artifacts required/available for performing the activity. Additionally, an activity often has social context represented by the element “community”. Interactions of the members of community with the subject are mediated by a set of rules governing the engagement of the members of community. On the other hand, the community interacts with the object through their division of labor towards achieving the object.

All the elements above constitute an “activity system” (Figure 1). The model depicts the idea that performance of an activity requires an adequate activity system, which can enact the “transformation process” that brings about change of state of the object in order to produce the outcome(s). If we try to understand a knowledge-work using this model, we can identify various elements involved in a knowledge-work with those of the model of activity. The correspondence is shown in Figure 2.

Also, AT provides us with the notion of hierarchical levels of activity that includes activity, action and operation. Thus and activity is performed as a chain of actions in order to achieve some objective or motive. An action, in turn, is a conscious, goal-directed execution of a chain of operations. Operations are well-defined routines that can be executed without worker's consciousness of underlying details. Identification of an activity with these levels is conditional upon the sophistication of the activity system.

For example, consider the situation when a researcher needs to access a paper. In a typical IT-enabled work environment, the researcher needs to launch a suitable application, formulate and fire a query, gets the link to the paper and downloads it. We can easily recognize this as an activity at the level of action. However, consider the environment where the paper can be found only in a physical library at the other end of the city. Getting the paper then involves definitely a significant activity. On the other hand, consider the other extreme, where the researcher utters the name of the paper and the system locates, retrieves and opens it on her computer screen. Here the activity is reduced to an operation from the worker's perspective.

## 3.2 Modeling a knowledge-work

Based on the theoretical grounding provided by AT, we formally model a knowledge-work as a tuple $a_i = \langle E_i, P_i, O_i \rangle$, where, $E_i$ is the set of entities, more specifically information about entities, their attributes and values (at performance-time) involved/required in performance of the activity. These entities are identified according to their roles with the elements of activity system described above (Fig 2) as their functional categories. $P_i$ represents the transformation process or simply process and $O_i$ represents the outcome(s) of the activity and nature(s) of the informational artifact for representing them.

To understand the above, let us consider the patient-care activity of diagnosis. Here, $E_i$ is consisted of the physician as the actor with attributes qualification, experience, etc.; results of the patient's examination and tests as well as a list of possible diseases constitute the tools; and the disease(s) to be confirmed forms the object. The informational outcome $O_i$ of the activity represent a list (initially empty, to be filled at performance-time) containing one or more diseases as the result of diagnosis. In this paper, for the sake of simplicity, we are not considering the community explicitly. However, even in this single-actor model, engagement of the community (e.g., hospital staff for patient-care) can be accommodated through special activities such as delegation, assignment, consultation, collaboration, etc.

With respect to the nature of $P_i$, we face two different possibilities. The first one is when $a_i$ is a cognitively manageable activity that will be called a "simple" activity. In a simple activity, a worker *can cognitively maintain the context and easily choose a chain of actions to be performed in order to reach the objective*. For example, recording of a patient's symptoms is a simple activity. The other possibility is that $a_i$ is sufficiently "complex" so that it needs to be decomposed into a number of sub-activities, each of which, in turn, may be $a_i$ complex one or simple one. We shall refer to such activities as "composite activities". The general structure of a composite activity is depicted in Figure 3.

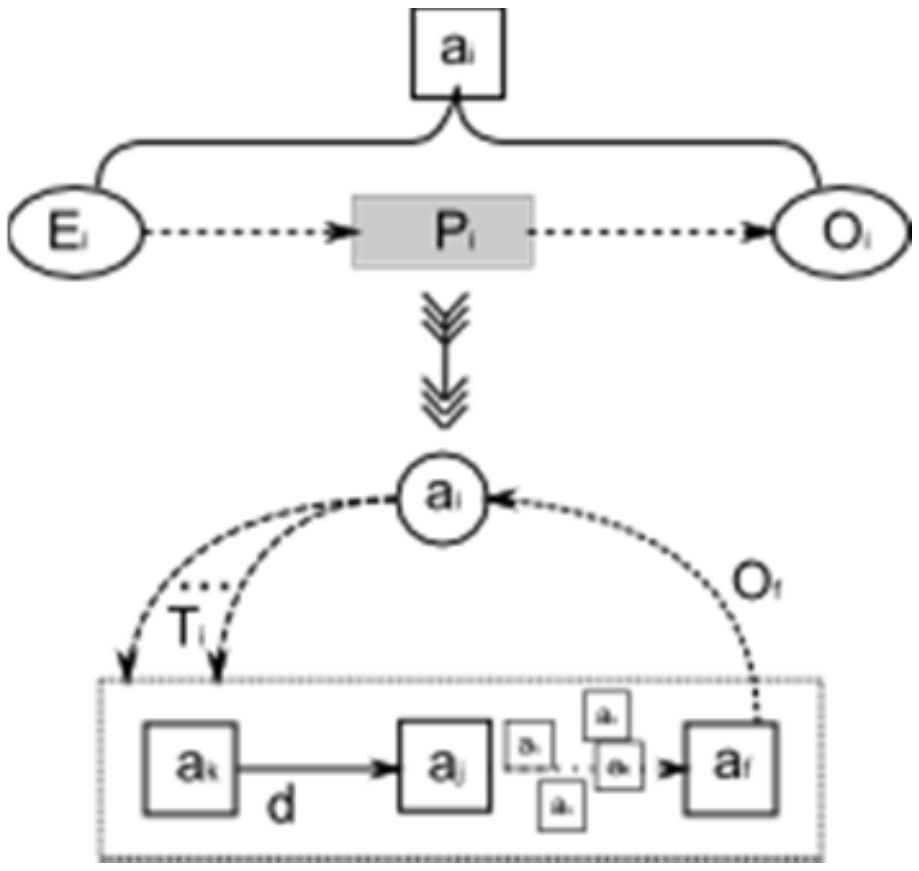

**Figure 3: Structure of a composite activity**

For example, to make a complicated diagnosis, based on results of examination, the physician may require hypothesizing a number of possible diseases, finding and recommending a set of clinical tests that will enable her to confirm or eliminate the possibilities. Once the test results are available, she needs to identify, in light of the test results, most probable disease(s) for which the patient needs to be treated. This complexity makes the activity of "diagnosing" as a whole a composite activity.

### *3.2.1 Structural properties of a composite activity*

For a composite activity $a_i$, the process $P_i = (V(P_i), E(P_i))$ is represented by a graph whose nodes $V(P_i)$ represent the set of sub-activities of $a_i$ and edges $E(P_i)$ represent their interrelationships. $P_i$ has the following properties:

- An activity $a_j \in V(P_i)$) inherits the "tools", $T_i$ of $a_i$, i.e., $T_j \cap T_i \neq \phi$;
- If for an activity $a_j \in V(P_i)$, $T_j \supseteq \{O_k\}$, where $\{a_k\} \in V(P_i)$, then $\{a_k\}$ is called the *support set* of $a_j$ and denoted as $SSet(a_j)$. The set of edges $\{e_{kj}\}$ are called the dependency edges or *d-edges* of $a_j$ (Fig. 3). Clearly, performance of $a_j$ cannot be started till performance of $SSet(a_j)$ is completed;
- There is *at least one* sub-activity $a_j \in V(P_i)$ for which $SSet(a_j) = \phi$. Performance of $a_i$ can be initiated with performance of any such sub-activity, denoted as the set $Init(a_i)$;
- There is *one and only one* sub-activity $a_f \in V(P_i)$ , called the *final (sub-)activity,* for which $O_f = O_i$ and whose completion denotes of the completion of the activity $a_i$.

### *3.2.2 Modeling a "target task-type"*

For building a KwSS that supports performance of episodes of a particular type of knowledge-intensive task, a typological or categorical model of the task needs to be built. This serves as the "reference" or "nominal" task-model for the KwSS. The modeling starts with consideration of the whole task as the largest unit of complex activity, $T = \langle E_T, P_T, O_T \rangle$, to be supported. Then, either based on a careful analysis or a recognized best practice, the elements of $E_T$ and $O_T$ as well as the structure of $O_T$ is identified. Then the elements of $E_T$ and $O_T$ are associated with their respective functional and semantic (may be drawn from a domain ontology, more on this later) typologies. Then $P_T$ is decomposed into its constituent sub-activities. The process is carried on recursively till the model includes all cognitively manageable simple activities required to be performed.

The resulting model serves as a standard or reference for all the episodes of performance of the target. Note that, the typologies in the reference model get bound to specific, episodic values during the performance of a task-episode. To understand this crucial point, let us consider the activity of diagnosing a patient. Its typological model carries the in-formation that its (1) tools are comprised of (informational) entities such as examination results (symptoms, signs, and medical history), possibilities considered, tests recommended and their results for a patient; (2) actor is a physician; and (3) object is a set of disease entities. However, the values to be bound with the entities, e.g., identity of the patient and the physician treating her, results of examination conducted on her, etc., are available only during the particular episode of treating the patient. The activity diagnosis is performed by the physician in order to find the values of the entities diseases for the patient-under-treatment.

# 4. AN ARCHITECTURE FOR KWSS

In this paper, we propose a software-agent based architecture of a KwSS as depicted in Figure 4. In the following we describe its components.

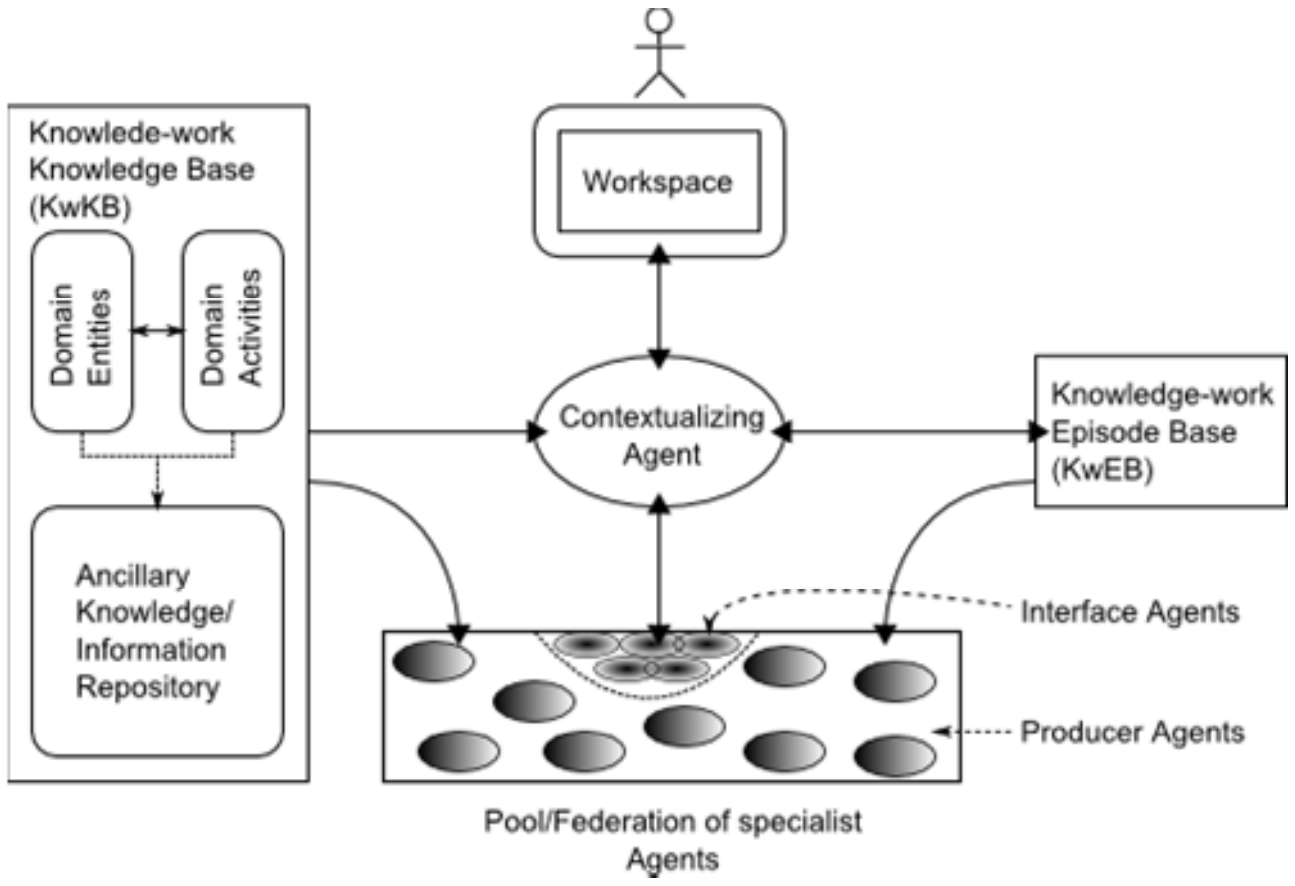

**Figure 4: Agent-based architecture of KwSS**

## 4.1 The Knowledge-work Knowledge Base

The "Knowledge-work Knowledge Base or KwKB" shown in Figure 4, is common sharable information repository for a KwSS. The "domain activities" component contains the reference task model described above for the target while the "domain entities" module contains one or more formal taxonomy/ontology of the terms and concepts relevant for the activities. Examples of such taxonomies include general ones such as Wordnet, Cyc etc. as well as domain-specific ones such as GALEN, UMLS, SNOMED-CT, etc., in biomedical domain [2]. Elements of a task-model are associated with relevant elements of these taxonomies in order to provide "semantics" to the activities and its constituent elements. This arrangement allows components of the KwSS system to carry out various types of semantically augmented computations about activities, information and their contexts.

A KwKB may also include one or more ancillary repositories of descriptive/explanatory information about domain activities and entities. The contents of these repositories, associated/indexed with the elements of the activity models and entities, can be a very useful resource for on-the-job learning by the knowledge-workers.

## 4.2 The Knowledge-work Episode Base

The other common and shared information repository in a KwSS is the "Knowledge-work Episode Base" or KwEB (Figure 4). The KwEB contains information about the instances or episodes of the target task-type performed by knowledge-workers using the KwSS. The logical organization of an episode-information adheres to the activity modeling formalism described earlier. However, contents of its activity elements are the "values" of the entities as they are determined/evaluated or articulated by the knowledge-worker within the scope of the particular episode. Also, the dependency structure of the activities and information reflects the "actual performance" of the task episode. Further, the episode information is tagged with the functional categories and semantic categories drawn from the task-model and entity taxonomies in KwKB (except some cases of "exercise of discretion" by the knowledge-worker, when the activity and entity typologies may be unavailable in KwKB).

Finally, at the level of simple activity, the episode is captured in terms of a chain of actions, where, each action refers to creation, evaluation, modification and various types of value-additions (e.g., adding reference/support) of relevant entities. Formally, the chain of action, leading to the production of the outcome entity of *e-th* episode of a simple activity, $a_i^e$, the $O_i^e$ is represented in KwEB as depicted in Figure 5. Overall, the KwEB is designed to support efficiently various types of computations, e.g., retrieval and navigation of relevant granular information from past episodes, to access information contents based on not only keywords, but also their context (including argumentative structure, provenance and lineage) and semantics.

$$
\left[
\begin{array}{lcl}
a_i^e :< E_i^e(0), P_i^e, O_i > & & \\
\hline
\quad P_i^e & & \\
\quad \downarrow & & \\
Act_1(E_i^e(0) : \{e_j\}_1 = \emptyset) & \rightarrow & \{e_j\}_1 \subset E_i^e(0) \\
 & & Op_1(\{e_j\}_1) \rightarrow e_1^* \\
E_i^e(1) \leftarrow e_1^* \cup E_i^e(0) & & \\
Act_2(E_i^e(1) : \{e_j\}_2 = \emptyset) & \rightarrow & \{e_j\}_2 \subset E_i^e(1) \\
 & & Op_2(\{e_j\}_2) \rightarrow e_2^* \\
E_i^e(2) \leftarrow e_2^* \cup E_i^e(1) & & \\
\vdots & & \vdots \\
Act_g(E_i^e(g-1) : \{e_j\}_n = \emptyset) & \rightarrow & \{e_j\}_n \subset E_i^e(n-1) \\
 & & Op_n(\{e_j\}_n) \rightarrow e_n^* \\
E_i^e(n) \leftarrow e_n^* \cup E_i^e(n-1) & & \\
\quad \downarrow & & \\
\hline
O_i^e \subseteq \{e_j^* \in E_i^e(n) | j = 1, \ldots, n\} & &
\end{array}
\right]
$$

**Figure 5: A chain of actions representing a simple activity**

## 4.3 Workspace

The "workspace" shown in Figure 4 is the interface of a KwSS that a knowledge-worker interacts with (directly or indirectly) in course of performing an episode $a_i^e$ of an activity at a time. If the activity is "composite" one, i.e., it is constituted of smaller sub-activities, the workspace presents her with the reference structure of activity and she chooses one of the permissible sub-activities. Here the term "permissible" refers to all sub-activities whose requisite initial entity set $E_i^e(0)$ is already value-bound. The chosen activity can be one recommended by the reference model or a new discretionary activity introduced by the worker, for which no or partial typology is available in KwKB. (Here we will not go into the details of handling discretionary choices.) The process may be repeated till the knowledge-worker reaches a cognitively manageable granular activity, i.e., simple activity. Once a simple activity to be performed is selected, she exercises her cognitive faculties in conjunction with available system support for performing the simple activity $a_i^e$ in order to achieve a set of objective(s) or goal(s). The nature of support provided by the workspace is described below.

At the beginning of performance of $a_i^e$, the workspace nominally presents to the user information about the set of relevant entities $E_i^e(0)$ along with their episode-specific values as well as a set of yet unvalued, i.e., typological or categorical entities $O_i^e$ constituting the objective or goal entities. The performance of $a_i^e$ is performance of a chain of actions that conclude successfully when the worker is able to assign relevant episodic values to the categorical goal entities.

Formally, the *t-th* action by the worker involves selecting a set of entities $\{e_j\} \in E_i^e(t-1)$ and apply an operation on them for computing a new entity (or the value/instance of an existing categorical entity) $e_j^*$ to the workspace so that the entity set is transformed into $E_i^e(t) \leftarrow e_j^* \cup E_i^e(t-1)$, as depicted in Figure 5. Clearly, there could be a large variety of entities and possible operations on them which may need to be accommodated in the workspace. In the following we divide them into two groups.

**Group 1** The operations involve entities either already available within the workspace, i.e., $E_i^e(t-1)$, or retrievable from internal repositories, the KwSS knowledge-base (KwKB) and episode-base (KwEB):

- assigning values to yet unvalued entities;
- arithmetic and logical operations on the entities with quantitative values;
- compilation of list of entities;
- creation of textual information entities (articulation) such as annotation, interpretation, summarization, analysis, conclusion, etc.;
- creation of referential links among entities in the workspace;
- retrieving granular information from past task-episodes;
- seeking details, including that of relationship information about an entity from KwKB, including its ancillary portion as we as episodic contextual information from KwEB; and
- semantic comparisons of entities – comparison of values (including textual) of semantically commensurate features of the entities, e.g., comparing symptoms and signs of a patient with those for a possible disease;

**Group 2** The class of operations designed for sending out and bringing in information to/from the external environment to the workspace:

- creation of contextualized "query package" from $E_i^e(t-1)$, exporting them to external IR or other systems/ services and importing their results to workspace; and
- Creation of contextualized "information package" from $E_i^e(t-1)$, exporting them to other actors by various means of communication for collaborative work and importing their results to workspace;

## 4.4 The Contextualizing Agent (CA)

The "Contextualizing Agent" or CA is at the heart of a KwSS. With the help of knowledge about the activities and entities drawn from the Knowledge-work Knowledge Base (KwKB), it tracks/monitors the progress of a knowledge-worker's performance in workspace. Continuous sensing of the work-context allows the CA (1) to guide the knowledge-worker through the maze of activities; (2) to locate and make available relevant resources (information about entities, people etc.), tools (including external ones) and other artifacts (e.g., computational protocols/templates)) in the workspace; (3) to ensure integrity in case of exercise of discretion by the worker; and (4) to capture, organize and archive new information created in the workspace into the episode base for future reuse. The functionalities of CA described above can be further enhanced and/or expanded in scopes though use of appropriate services provided by one or more "specialist agents" (SA) from the agent pool shown in Figure 4. We shall discuss them in section 4.5.

### *4.4.1 Reasoning with activities*

During the performance of an episode of an activity, the contextualizing agent or CA needs to decide on the nature (e.g., guidance, action-level supports) and content (e.g., information, computation template) of the support it provides at a point of time. The problem needs the CA to "reason" about an activity and its state of progress. The reasoning process encompasses the typological models available in the knowledge base and the current states and informational contents of the episodic performance of the current activity and other activities related to it. Figure 6 depicts an algorithmic description of the core reasoning process followed by the workspace.

As indicated in algorithm 1, at any given time during an episode, the CA maintains three disjoint sets of activities, (1) the *ActiveSet* comprised of all activities currently being performed; (2) the *ReadySet* containing all activities whose performance can be started; and (3) *CompleteSet* consisting of all completed activities. At a given moment, there can be a number of activities in the activity model outside these sets. However, as the performance progresses, each activity moves through these sets so that at the end of performance ActiveSet and ReadySet are empty and CompleteSet contains all the activities.

### *4.4.2 Action-level supports*

The reasoning leading to action-level support corresponds to the block 4b of algorithm 1. However, no details are provided regarding their possible realizations. The first type of support is related to locating information resources from the internal sources, namely, the knowledge base and the case base. This is formulated as a problem of Case-Based Reasoning (CBR) [10]. To provide this support as the operation involved in *t-th* action, the CA constructs a probe/query vector drawing typologies and values of entities in current entity set $E_i^e(t-1)$. The selection and retrieval of information is performed based on the similarity of the probe vector (1) with the context of their production as recorded in the case base; and (2) with the typological contexts of the elements of knowledge base.

The support type described in item 2 of block 4b depends on the availability of suitable interfaces and/or communication channels to external resources. The CA routes suitably contextualized (by user and/or CA itself) information and/or query packages to external resources available to KwSS and channels their response to the workspace. The next support type in block 4b involves the

**Algorithm** 1. Support Performance($a_i^e$)

1. *ActiveSet* $\leftarrow a_i$;
2. *Retrieve the typological model* $a_i$ *from KB;*
3. *If* $a_i$ *is a composite activity then*
   (a) $\forall a_j \in Init(a_i) \subset V(P_i), ReadySet \leftarrow a_j$;
   (b) $a_k \leftarrow$ Guide user(*ReadySet*);
   (c) Support Performance($a_k^e$);
   (d) *If* $a_k = a_f \in V(P_i)$ *is available,*
      i. $O_i^e \leftarrow O_k^e$;
      ii. *ActiveSet*($\leftarrow$ *ActiveSet* $- a_k$)
      iii. *CompleteSet* $\leftarrow a_i$;
      iv. *Capture and archive* $a_i^e$ *and its dependency relationships;*
      v. *Exit;*
   (e) *Else*
      i. Update ReadySet($a_k$)
      ii. $a_k \leftarrow$ Guide user(*ReadySet*);
      iii. Support Performance($a_k^e$);
4. *Else if* $a_i$ *is a simple activity then*
   (a) *Present to the user through* ***Workspace***

$$T_i^e \leftarrow T_{i'}^e \bigcup_{a_j \in SSet(a_i)} O_j^e,$$

   *where* $a_{i'}$ *is the super-activity of* $a_i$;
   (b) *Track user's progress through a chain of actions in the workspace and provide on-demand as well as pro-active action-level supports such as:*
      - *Providing context-aware access to relevant granular information from knowledge base (including ancillary part) and case base;*
      - *Mediating access to external tools and information sources for implementation of operations in workspace requiring them ;*
      - *Identifying and mediating access to the services of one or more pooled SA for more sophisticated supports such as recommendation, context-aware search etc.;*
      - *Managing "discretion" exercised by the user;*
   (c) *Sense the completion of activity, successful or otherwise;*
   (d) *If not successful,* Manage Failure;
   (e) *Otherwise*
      i. *ActiveSet* $\leftarrow$ (*ActiveSet* $- a_i$)
      ii. *CompleteSet* $\leftarrow a_i$;
      iii. *Capture information about* $a_i^e$ *and archive them in the* Episode/Case Base *in such a manner that the structure and contents of its performance can be reconstructed.;*
      iv. *Exit;*

**Algorithm** 2. Guide user(*ReadySet*);

1. *Present ReadySet within workspace for user to choose from;*
2. *Return user's choice* $a_k \in ReadySet$, *the next activity to perform;*

**Algorithm** 3. Update ReadySet($a_j$);

1. *Identify* $\{a_k\} \subset V(P_i)$ *such that* $a_j \in SSet(a_k)$;
2. *ReadySet* $\leftarrow \{a_l \in \{a_k\} | SSet(a_l) \subset CompleteSet\}$;

**Figure 6: Algorithm (s) for activity reasoning**

CA anticipating user's information needs and suitably orchestrating services available from the specialized agents (SA) to meet them. We will discuss about the SA in the next section.

A user's discretion (item 4 in block 4b) is essentially results in a user-introduced deviation from the nominal activity structure indicated by the typological model. Such deviations include skipping/deleting an activity, introducing new activity and substitution of a simple activity with a composite activity and vice versa. The CA allows them at the episode level and ensures that the integrity (e.g., the support set of another activity should not become empty due to deletion of an activity) of the activity structure (section 3.2.1) is maintained. If such deviations include introduction of novel activity, it may not be possible for the CA to categorize its elements in episode base with typologies from KwKB. In such a case, the CA might be configured to encourage the worker to provide relevant typological information.

In an activity episode, if the worker is unable to find pertinent values of the goal entities, the performance is perceived as a failure (line 4d). The cause of the failure may be rooted in the inadequate performance of an earlier activity. For example, while diagnosing a physician may fail to reach a firm diagnosis because during the examination some signs or symptoms were overlooked. To remedy the failure the worker needs to re-perform a number of activities, typically starting from the cause activity up to the failed activity. However, during the re-performance, the information produced earlier through those activities are also available to the worker. Through failure management the CA facilitates the above in a manner consistent with the structural properties of the affected activities.

## 4.5 The Pool of Specialist Agents

The CA in a KwSS is designed to provide a set of essential/nominal supports. It is easy to envisage a wide range of enhancements as well as extensions of scopes of these nominal capabilities with the aid of various cutting-edge computational techniques. These techniques are emerging in various areas of research such as text analysis, semantic categorization and reasoning, information retrieval etc. Currently they are in a state of rapid and continuous evolution. Naturally, their effectiveness in a KwSS will also evolve as it accommodates the gradual progress made in these fields.

However, such accommodations and their management pose difficult design challenges. If they are not carefully insulated from the core of the system, they can easily destabilize it, even at the level of its nominal functionalities.

In view of the above concerns, in proposed agent-based architecture of a KwSS, these enhancements are implemented as services from a pool of specialist agents (SA). The pool is an ecosystem of collaborative agents with varying degrees of autonomy. As depicted in Figure 4, they fall into two categories, the interface agents (IA) and the producer agents (PA). An IA provides a higher level service, such as context-aware information retrieval, recommendation (about relevant information, computation protocol/template, external tools), resource (people/experts, artifacts) location etc. The CA, and in turn, the user in workspace are aware of these services. The CA can invoke these services on-demand from the worker. Also these agents can be selectively configured to provide their services pro-actively in response to the work-context as maintained by the CA.

An IA essentially provides its service as one or more (alternative) compositions of the services of a set of agents that can include relevant producer agents and other interface agents. A producer agent or PA has a specific, narrowly defined capability and well-defined service protocol to invoke the capability. The service of a PA can be used by more than one IA as part of their respective composition. For example, consider a PA whose service is to detect key entities and their relationships from a passage of natural language text. This is a vital service for an IA that tries to identify, based on the work-context, relevant text passages from a large document. The service of the same PA may be utilized by an IA for context-aware IR, for analyzing a natural language query.

This architectural approach allows us to continuously enhance the capabilities of a KwSS by improving quality of services of one or more specialist agents, at a time, without disrupting other services or functionalities of CA. Such improvements can be brought about by modifying the techniques/algorithms they employ, suitable reinforcements of their specialized KB etc. In our works with KwSS, we are working on design and development of a number of specialist agents. Most exciting among them is what we call the "generative resource modeling agent". It identifies various resources (information, computational protocols) and profiles their scope(s) of utility through inductive and/or abductive analysis of episode base. It can enable a KwSS to adaptively "chunk" and "operationalize" some parts of chains-of-actions based on evolving patterns among the historical work-contexts.

## 5. CONCLUSION

Building a KwSS, as proposed in [9] for providing comprehensive task-specific support to knowledge-workers, requires multi-faceted effort in computing research and engineering. In identifying and solving relevant problems we may need to co-opt ideas and concepts from a various fields outside computing/IT research and practices. In this paper we have attempted to develop some important building blocks, namely, a software-agent based architecture for knowledge-based computing, an activity modeling technique, a formal representation of activity models in a knowledge-base and an approach for reasoning with these models. We believe that these will contribute significantly in building KwSS as well other KwSS-like systems. Other important facets of the problem include design of suitable user interfaces in workspace that can allow efficient and intuitive presentation of large volume of information, collaborative connectivity across multiple platforms like mobile devices, integration with productivity/office applications, etc. On the methodological side, there is a need for developing suitable methodologies to collect and analyze information on target task(s) that can be translated into robust activity models. In our lab we have implemented an early version of a web-based patient-care KwSS. We are currently investigating some of these issues for enhancing the prototype.